\documentclass[twocolumn,preprintnumbers,amsmath,amssymb]{revtex4}
\usepackage{graphicx}
\usepackage{dcolumn}
\usepackage{bm}
\begin{document}
\title{\bf Thermomechanical properties of graphene: valence force field model approach}
\author{A. Lajevardipour$^1$, M. Neek-Amal$^{2}$ and F. M. Peeters$^3$ }
\affiliation{
$^1$Department of Physics, Eastern Mediterranean University, G. Magusa, North Cyprus, Mersin 10, Turkey.\\
$^2$Department of Physics, Shahid Rajaee Teacher Training University,
Lavizan, Tehran 16785-136, Iran.\\$^3$Departement Fysica,
Universiteit Antwerpen, Groenenborgerlaan 171, B-2020 Antwerpen,
 Belgium.}
\date{\today}

\begin{abstract}
Using the valence force field model of Perebeinos and  Tersoff
[Phys. Rev. B {\bf79}, 241409(R) (2009)], different energy modes of
suspended graphene subjected to tensile or compressive strain are
studied. By carrying out Monte Carlo simulations it is found that:
i) only for small strains ($|\varepsilon| \lessapprox 0.02$) the
total energy is symmetrical in the strain, while it behaves
completely different beyond this threshold; ii) the important energy
contributions in  stretching experiments are stretching, angle
bending, \textbf{out-of-plane term and a term that provides
repulsion against} $\pi-\pi$ misalignment; iii) in compressing
experiments the two latter terms increase rapidly and beyond the
buckling transition stretching and bending energies are found to be
constant; iv) from stretching-compressing simulations we calculated
the Young modulus at room temperature 350$\pm3.15$\,N/m, which is in
good agreement with experimental results (340$\pm50$\,N/m) and with
ab-initio results [322-353]\,N/m; v) molar heat capacity is
estimated to be 24.64\,J/mol$^{-1}$K$^{-1}$ which is comparable with
the Dulong-Petit value, i.e. 24.94\,J/mol$^{-1}$K$^{-1}$ and is
almost independent of the strain; vi) non-linear scaling properties
 are obtained from height-height correlations at finite temperature; vii) the used valence force field model
 results  in a temperature independent bending modulus for graphene, and viii)\textbf{ the Gruneisen parameter
 is estimated to be 0.64.}
\end{abstract}

\maketitle

\section{Introduction}
Since the discovery of graphene in 2004, which is an almost two
dimensional crystalline material, its exceptional mechanical
properties have been
studied~\cite{Lee,Giem2008,mayer,fasolinonature,arxiv2010,naturenanotechnology}.
Tensional strain in monolayer graphene affects its electronic
structure. For example strains larger than $15\%$ changes graphene's
band structure and leads to the opening of an electronic
gap~\cite{naturephys}.  In recent experiments the buckling strain of
a graphene sheet that was positioned on top of a substrate was found
to be six orders of magnitude larger (i.e. $0.5-0.6\%$) than for
graphene suspended in air~\cite{arxiv2010}. Furthermore, some
experiments showed that a compressed rectangular monolayer of graphene on
a plastic beam with size 30$\times$100 $\mu m^2$ is buckled at about
0.7$\%$ strain~\cite{compressionamall}.

 Elasticity theory for a thin
continuum plate and the empirical interatomic potentials (EP) are
two main theoretical approaches that have been used to study various
mechanical properties measured in compressing and stretching
experiments~{\cite{fasolinonature,fasolino,neekprb82}. Continuum
elasticity theory does not give the atomistic features of graphene
while the EPs, such as the Brenner potential
(REBO)~\cite{brenner,brenner2002} and the LCBOPII
potential~\cite{LCBOP}, can properly account for the mechanical
properties of graphene. Despite the several benefits of these EPs,
some special atomistic features of graphene subjected to compressive
or tensile strains could not be explained. The different energy
contributions  in these potentials are mixed. For example, in  REBO
all the many body effects  are put in the bond order term and the
different important energy  contributions are not separable.

It still remains  unclear  how large are the contributions of the
different energy terms in strained graphene. Using the recently introduced
\emph{valence force field} (VFF) model by Perebeinos and
Tersoff~\cite{Tersoff2009} we show how  the contribution of the different
energy modes in strained graphene can be separated and we calculate their
dependence on the value of strain.

The bending modulus of graphene at zero temperature was estimated
using several interatomic potentials, e.g.  the first version of the
Brenner potential ~\cite{brenner} yields 0.83~eV, the second
generation of the Brenner potential~\cite{brenner2002} estimated it
to be 0.69~eV, adding the third nearest neighbors (the dihedral
angle effect) in the Brenner potential enhances it to
1.4~eV~\cite{lu2009}, using the LCBOPII potential and continuum
membrane theory the bending rigidity was found to be
 0.82 eV~\cite{fasolinonature,LCBOP}, Tersoff's VFF model estimated it to be 2.1~eV,
and from ab-initio energy calculations it was found to be 1.5~eV~\cite{Yakobson} (note that
 `bending rigidity' (`bending modulus') is used for a membrane stiffness (an atomistic sheet)).
  Despite these  studies the temperature dependence of the bending modulus is poorly known.
   An increasing behavior versus temperature for the bending rigidity  was found ~\cite{fasolinonature} by using Monte Carlo simulations
   with the   LCBOPII potential and membrane
  theory concepts.  In contrast, Liu~\emph{et al}~\cite{liuAPL} found a decreasing
  bending rigidity with temperature using the REBO. Here we show that the  VFF model  predicts
  a temperature independent bending modulus.

In this study we employ VFF and carry out standard Monte Carlo
simulations in order to  calculate and compare the different energy
modes of a graphene sheet that is subject to axial strains. The
total energy is found to be different for compressing and stretching
when strains are applied larger than $|2|\%$. Two important terms,
i.e. stretching and bending, vary differently depending on the way
that one stretches or compresses the system. We find that
out-of-plane and $\pi-\pi$ terms have much larger contributions in
compression experiments when compared to stretching. Furthermore, we
used this potential to calculate  Young's modulus at room
temperature
 from stretching-compressing simulations. We also calculate the molar heat capacity.
Our Monte Carlo simulations show that the VFF potential yields
a temperature independent bending modulus.

This paper is organized as follows. Section II contains the essentials of the
VFF model for graphene. The simulation method for strained graphene will
be presented in Sec. III. Different energy modes of strained graphene are
 studied in Sec. IV. In Sec. V the molar heat capacity for
  non-strained suspended graphene is calculated. Temperature effects
  of the bending modulus of graphene with periodic boundary condition are presented in
  Sec. VI and the scaling properties of graphene at finite temperature are investigated in Sec VII. We will conclude the paper in Sec. VIII.

\section{Elastic energy of graphene}
There are two main classical approaches for the investigation of the elastic energy
of graphene: 1) the continuum approach based on elasticity theory,
and 2) the atomistic description using accurate interatomic potentials.

\textbf{The total energy of a deformed membrane consists of two
important terms: stretching and bending. For almost flat and
continuum membrane using Monge representation the surface area
element $dA$ can be approximated by a flat sheet area element in the
x-y-plane, i.e. $dA\approx dxdy$ and the bending energy is written
as $\frac{1}{2}\int dxdy \kappa (\nabla^2h)^2$ where $\kappa$ is the
bending rigidity and $h$ is the out-of-plane deformation of the
membrane at point $(x,y)$. The stretching term for an isotropic
continuum material in the linear regime includes two independent
parameters: the shear modulus ($\mu$) and the Lam\'{e} coefficient
($\lambda$) and is written as $\frac{1}{2}\int dxdy [2\mu
u^2_{\alpha\beta}+\lambda u_{\alpha\alpha}^2]$. Here
$u_{\alpha\beta}=\frac{1}{2}[\partial_{\alpha}u_{\beta}+\partial_{\beta}u_{\alpha}+\partial_{\alpha}h\partial_{\beta}h]$
is the second rank symmetric tensor with $\alpha,\beta=1,2$ and
$u_{\alpha}(x,y)$ is the $\alpha^{th}$ component of the displacement
vector. Neglecting the last term in the strain tensor makes the
stretching term linear and decouples the bending and stretching
energy. Therefore for an isotropic and continuum material for small
deformations  and with the assumption of a nearly flat membrane
($|\nabla h|^2\ll 1$) the strain energy ($U_T$) can be written
as~[18]
\begin{equation}\label{Monge}
U_T=\frac{1}{2}\int dxdy [\kappa (\nabla^2h)^2+2\mu
u^2_{\alpha\beta}+\lambda u_{\alpha\alpha}^2].
\end{equation}
}
 The integral is taken over the projected area of the membrane into
the x-y-plane.
 For isotropic materials and in the linear approximation the
mentioned parameters are related to the Young modulus ($Y$) and
Poisson's ratio ($\nu$) as $\mu=Y/(2(1+\nu))$ and $\lambda=2\mu
\nu/(1-2\nu)$. Equation~(\ref{Monge}) can be rewritten in terms of
the Fourier components of $h$ and yields the scaling properties of
the sheet. Despite these benefits, this continuum model does not
include self-avoidance, the natural condition of true physical
systems and does not show  atomistic details of the membrane under
different boundary conditions. All these deficiencies originate from
the continuity assumption. Assuming graphene as a continuum plate
limits the study to only bending and stretching modes.

\textbf{Due to the hexagonal symmetry of the flat monolayer graphene
lattice, it is elastically isotropic which implies that the
 the bending modulus is independent of the direction at
least within the linear elastic regime \cite{lu2009}. However, the
graphene monolayer can exhibits anisotropic behavior in  the
nonlinear regime where distortions are no longer infinitesimal. The
larger stretches, the stronger anisotropy and non-linearity effects.
Cadelano \emph{et al} found that monolayer graphene is isotropic in
the linear regime, while it is anisotropic when nonlinear features
are taken into account~\cite{poisson}.}

~~~


The recently introduced VFF model in Ref.~\cite{Tersoff2009} is expected to be able to describe
both compression and stretching
experiments by separating the contribution of the various energy modes. This model
 includes explicitly the various relevant energy
terms which describe the change in the bond lengths, bond angles and
torsional effects. The total  energy density is written as

\begin{equation}\label{general}
E_T=\frac{1}{NS_0}(E_{st}+E_{be}+E_{out}+E_{bo}+E_p+E_{co}),
\end{equation}
where $N$ is the number of atoms and $S_0=\frac{3\sqrt{3}}{4}a_0^2$
is the surface area of the unit cell of the honeycomb lattice.
In the following we will discuss the different terms in Eq. (2).
Note that the energy
reference is set to zero. Assuming $a_0=1.42\,\AA$~as the unit of
length, the `\emph{stretching}' and `\emph{bending}' (bending of the
bond angle) terms are
\begin{eqnarray}
                  E_{st} &=&\frac{1}{2}K_s\sum_{i,j}(\delta r_{ij})^2 \\
                  E_{be} &=&K_{be}\sum_{i,j<k}(\cos(\theta_{ijk})-\cos(\theta_0))^2
 \end{eqnarray}
where $\delta r_{ij}=r_{ij}-1$ and $\theta_0=2\pi/3$. In Eqs. (3)
and (4) $r_{ij}$ is the bond length between atom `i' and `j',
$\theta_{ijk}$ is the angle between the nearest neighbor atoms `i',
`j' and `k' and $\theta_0$ is the equilibrium angle between three
nearest neighbor atoms. $E_{st}$ is the two-body stretching term
responsible for  bond stretching. $E_{be}$ is  the bending energy
due to the bond angles. Here, all bond angles will be considered.
The above two terms results in a quasi-harmonic model~\cite{LCBOP}.
Later, we will find that these two terms become constant as function
of strain when beyond the buckling point in a compression
experiment.

 The stiffness against
`\emph{out-of-plane}' vibration is provided by

\begin{equation}
E_{out}=K_{out}\sum_{i,j<k<l}(\frac{3\overrightarrow{r_{ij}}\cdot\overrightarrow{r_{ik}}\times
\overrightarrow{r_{il}}}{r_{ij}r_{ik}+r_{ij}r_{il}+r_{ik}r_{il}})^2,
\end{equation}
where the summation is taken over the first neighbors of atoms $`i$'
and taking care of not double counting. In Eq. (5) $\overrightarrow{r_{ij}}$ is the distance vector between atom `i'
and `j'. Hence there are three different terms for each atom. Correlations between bond lengths are provided
by the '\emph{bond order}' term
\begin{equation}
E_{bo}=K_{bo}\sum_{i,j<k} \delta r_{ij}\delta r_{ik},
\end{equation}
where for each bond length with central atom `i' three different terms
are considered.

The misalignment of the neighboring $\pi$ orbital is given by the
`\emph{$\pi-\pi$}' term
\begin{equation}
E_{p}=\frac{1}{2}K_{p}\sum_{i,j} |\overrightarrow{\pi_i}\times
\overrightarrow{\pi_j}|^2,
\end{equation}
where
\begin{equation}
\overrightarrow{\pi_i}=3\frac{\overrightarrow{n_{ijk}}+\overrightarrow{n_{ikl}}+\overrightarrow{n_{ilj}}}
{r_{ij}r_{ik}+r_{ij}r_{il}+r_{ik}r_{il}}. \end{equation}
$\overrightarrow{n_{ijk}}=\overrightarrow{r_{ij}}\times\overrightarrow{r_{ik}}$
is a vector normal to the plane passing through the vectors
$\overrightarrow{r_{ij}}$ and $\overrightarrow{r_{ik}}$. This kind
of interaction plays an important role in the interlayer interaction
in graphitic structures. Note that the simple two body interaction
gives only $2\%$ of the local density of state (LDA)  result for the
energy difference between AA and AB stacked graphite~\cite{prb2005}.

The last term takes into account the coupling
between bond stretching and bond angle bending (bond length-bond
angle cross coupling), i.e. the \emph{`correlation'} term
\begin{equation}
E_{cor}=K_{cor}\sum_{i,j<k}\delta
r_{ij}(\cos(\theta_{ijk})-\cos(\theta_0)).
\end{equation}

The coefficients in the above equations ($K_s,K_{be}$ and so on) were recently parameterized
by Perebeinos and Tersoff~\cite{Tersoff2009} such that the phonon dispersion of graphene was accurately described.
 These parameters
are listed in Table I.

\begin{table}
\caption{ Parameters of the energy model (Eq.~(\ref{general})) are taken from Ref.~\cite{Tersoff2009}. Units are in
eV.}
\begin{tabular}{|c|c|c|c|c|c|}
\hline
$K_s$&$K_{be}$&$K_{out}$&$K_{bo}$&$K_{p}$&$K_{cor}$\\
 \hline
37.04&4.087& 1.313& 4.004& 0.016102& 4.581\\
\hline
\end{tabular}
  \centering
\label{tablestretch}
\end{table}

\begin{table*}
\caption{Young's modulus of graphene in units of N/m.}
\begin{tabular}{ | c | c  c  c|  c  c  c  c c | c |}
 \hline
Experimental&& Classical (T=300\,K)&&&&Ab-initio (T=0\,K)&&&Tight-Binding\\
 \hline
 340$\pm$50&350$\pm$3.15&355$\pm$21&384&345$\pm$6.9&336&352.54&351.75&322&312\\
 \hline
Ref.~\cite{Lee}&Present$^{*}$&Ref.~\cite{fasolino}$^b$&Ref. ~\cite{TersoffBrenner}$^{c}$&Ref.~\cite{Yakobson}&Ref.~\cite{Ortwin}&Ref.~\cite{diamond}&Ref.~\cite{Liu}&Ref. \cite{Jphys}&Ref.~\cite{poisson}\\
 \hline
\end{tabular}
  \centering\\
$^{*}$ VFF model~\cite{Tersoff2009}, $^b$ LCBOPII potential, $^c$
Tersoff-Brenner potential
.\label{table2}
\end{table*}

\section{Simulation method: strained graphene}

In order to compress (stretch) graphene nanoribbons, we have carried
out several standard Monte Carlo simulations~\cite{fasolinonature}
of a suspended graphene sheet at finite temperature.
Equation~(\ref{general}) is used to calculate the total energy of
the system. Our sample is a rectangular graphene sheet with $l_x
\times l_y$ dimensions in $x$- and $y$-directions, respectively,
containing $N_0$=1600 atoms.  The sheet is strained along the
armchair or the zig-zag direction. Strain is always applied along
$x$. When  graphene is strained along the armchair direction we
named it \emph{armchair graphene} -AC-
($l_x$=85.2\,\AA,~$l_y$=49.19\,\AA) and  strained graphene along
zig-zag direction is named \emph{zig-zag graphene}-ZZ-
($l_x$=98.38\,\AA,~$l_y$=42.6\,\AA). Periodic boundary conditions
are applied along the lateral direction i.e. zigzag direction in  AC
graphene and along the armchair direction for ZZ graphene. Our
simulation starts with a flat sheet, and we allow then the system to
thermally equilibrate such that the total energy no longer changes.
Temperature is typically taken $T$=300\,K, except when otherwise
indicated.
 Figure~\ref{fig0}(a) shows a snap shot of the relaxed unstrained ZZ graphene at
$T$=300\,K (note that the supported ends
are fixed). We found that the graphene sheet is corrugated after
relaxation which are the intrinsic thermal ripples in graphene. Thus the
used VFF is able to display true structural properties. These ripples are
vital in order to make suspended graphene stable and are therefore crucial for
the stability of a flat 2D crystal at finite
temperature~\cite{fasolinonature}.

To simulate a suspended sheet we fixed two atomic rows at both longitudinal ends.
 These boundary atoms are not
included in the summations when calculating the
different energy contributions (in Eqs. (2-9)), i.e. $N=N_0-40$. We
compress/stretch the system with a slow rate, i.e. in every million
Monte Carlo steps the longitudinal ends are reduced/elongated with  about
$\delta=$0.02~\AA~ such that the system stays in thermal equilibrium.
After obtaining the total desired strain $\varepsilon$,
we wait for an extra 4 million steps during which the system can relax.
For example, a strain (applied in $x$-direction) of $\varepsilon=1.2\%$~is achieved after 29$\times 10^6$
Monte Carlo steps.

\begin{figure}
\begin{center}
\includegraphics[width=1\linewidth]{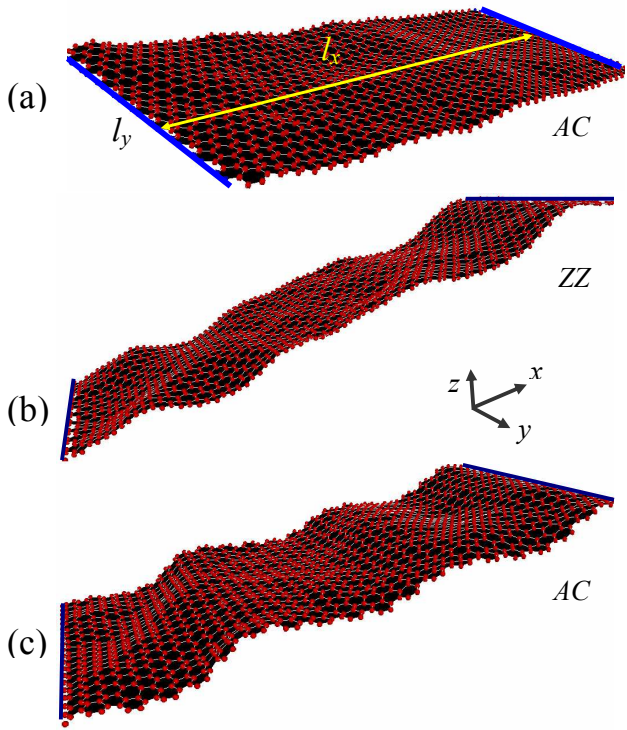}
\caption{(Color online)  Snap shot of a suspended graphene sheet
at $T$=300\,K using the valence force model (Eq.~(\ref{general})).
Blue lines indicate the position of fixed atoms in $x-y$ plane. (a) Unstrained, (b) compressed ZZ graphene, and (c) compressed AC
graphene. \label{fig0}}
\end{center}
\end{figure}

\begin{figure*}
\begin{center}
\includegraphics[width=0.3\linewidth]{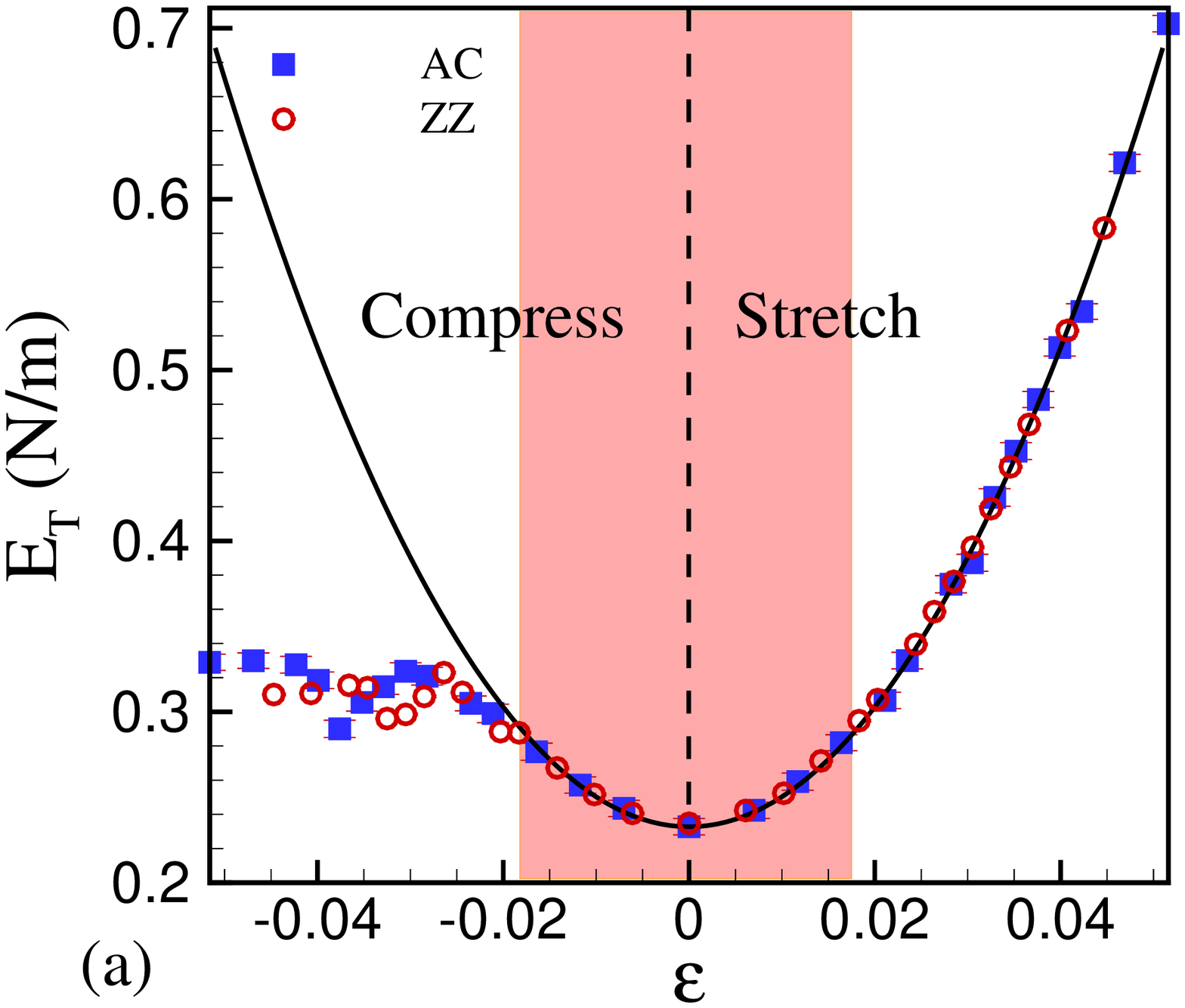}
\includegraphics[width=0.3\linewidth]{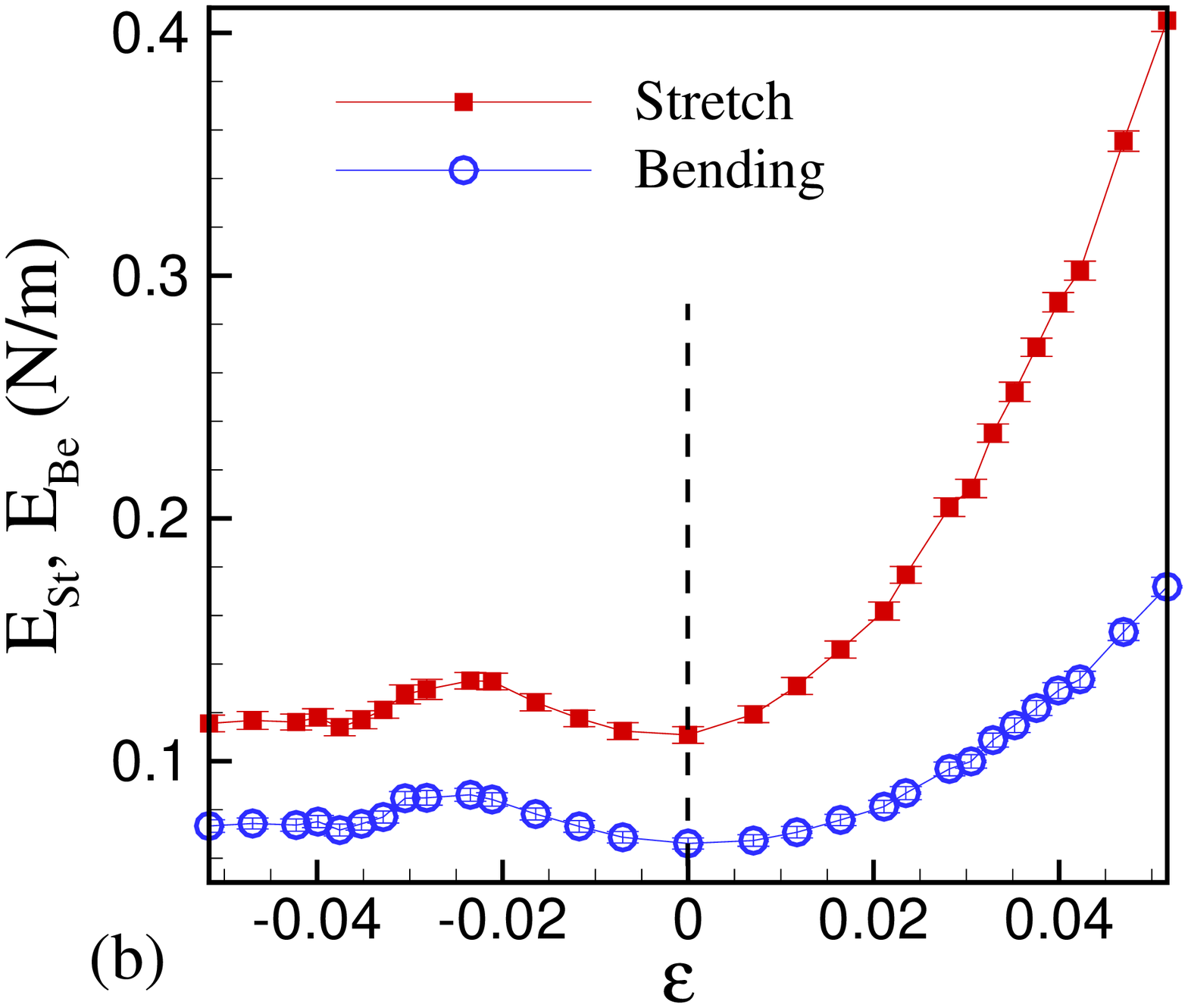}
\includegraphics[width=0.3\linewidth]{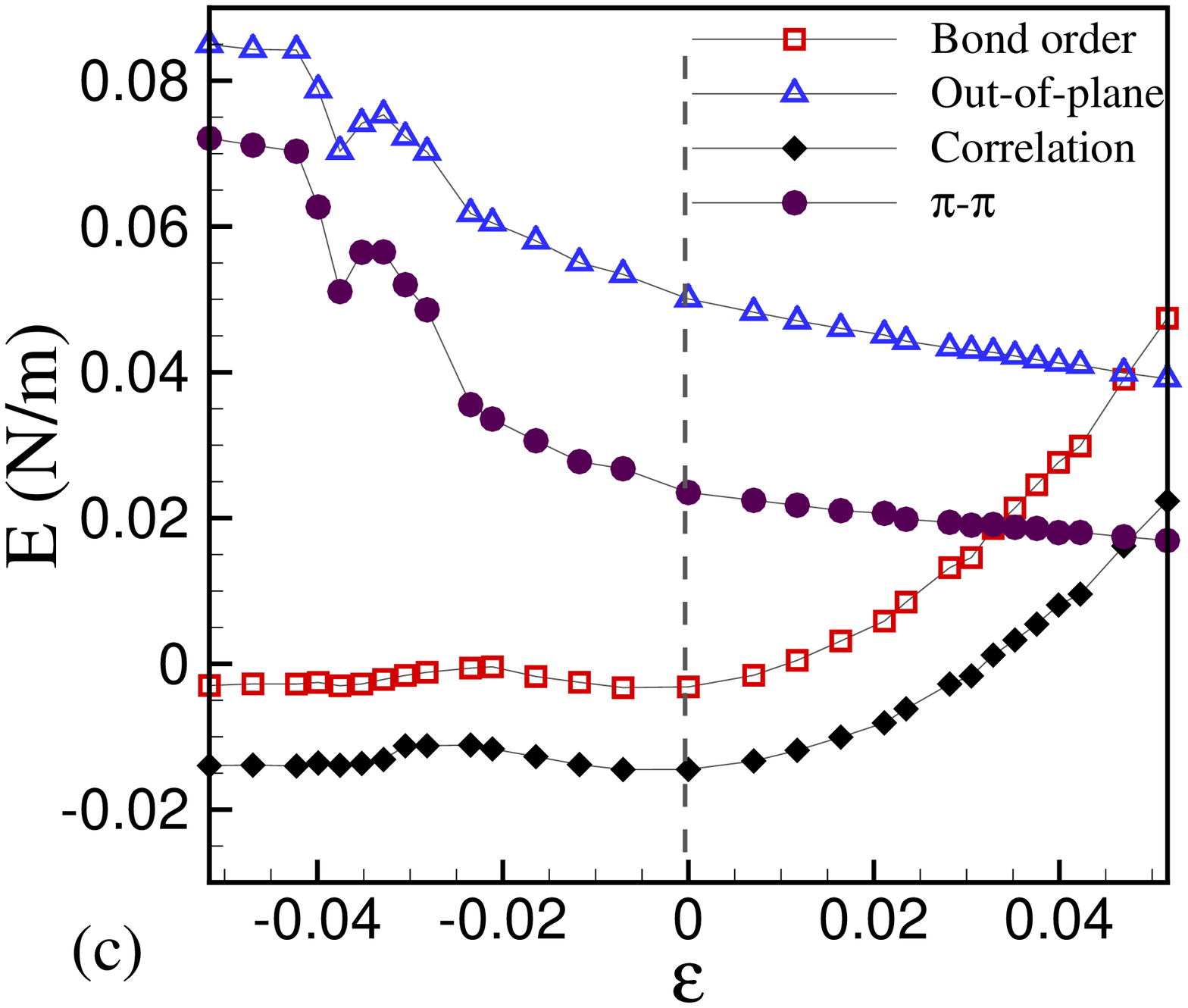}
\caption{(Color online) (a) Total energy of a graphene sheet subjected
to stretching and compression for AC and ZZ. (b) Contribution of the bending (Eq.~(4)) and the
stretching  (Eq.~(3)) terms of the total energy for AC. (c) Contribution of the other remaining terms
given by Eqs.~(5)-(9) for AC. \label{fig3} }
\end{center}
\end{figure*}

\begin{figure}
\begin{center}
\includegraphics[width=1.\linewidth]{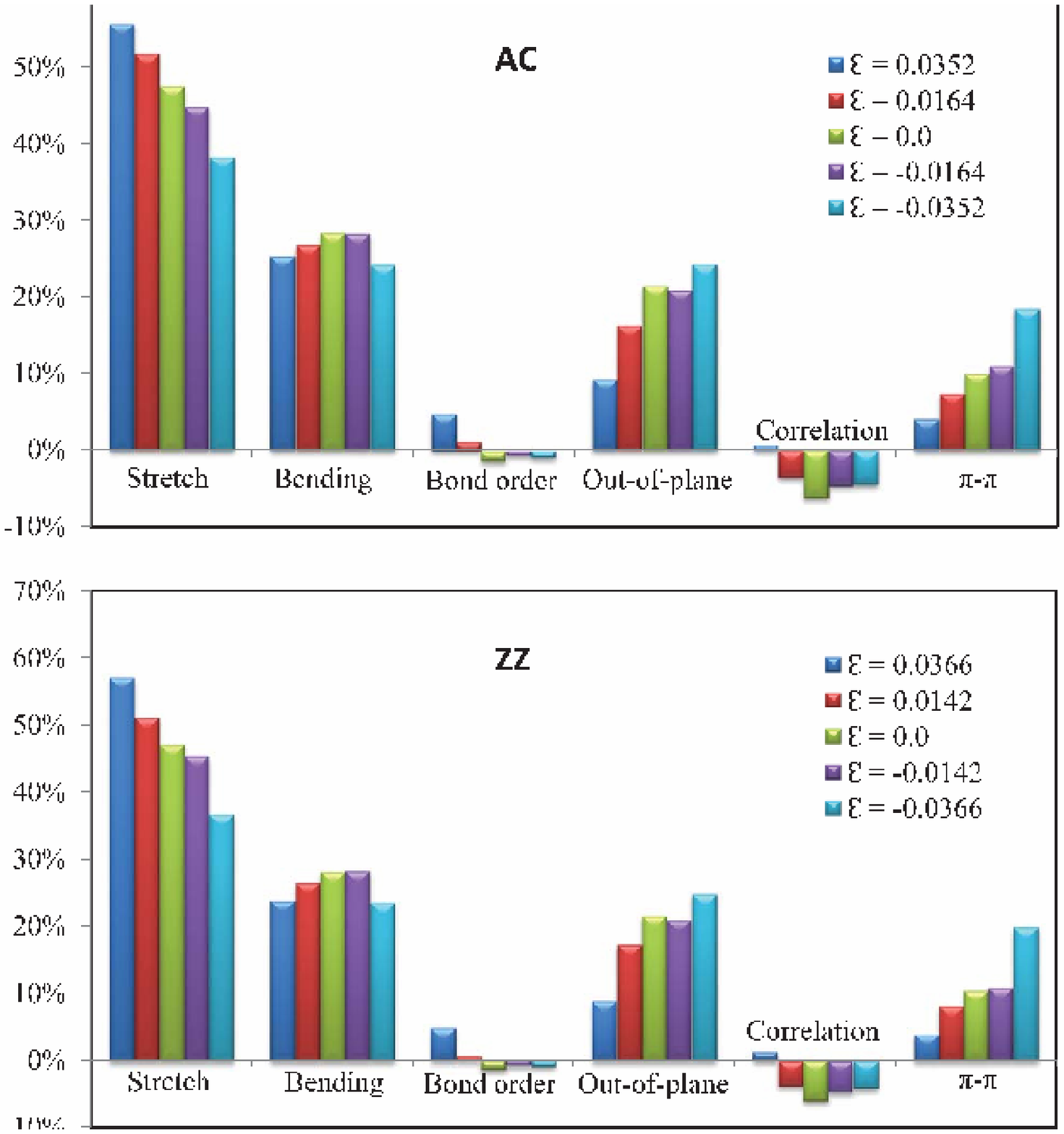}

\caption{(Color online) Contribution of the different energy terms to the total energy for three typical
values of the strain in AC (Top) and ZZ (Bottom) graphene. \label{figp} }
\end{center}
\end{figure}

\section{Different energy modes for strained graphene}
Figures~\ref{fig0}(b,c) show two snap shots of compressed ZZ and AC
graphene, respectively when $\varepsilon$=-2$\%$. It is interesting
that the rippled structure is different in the two cases. This is
due to the different out-of-plane and $\pi-\pi$ interaction terms
around and beyond the buckling transition points, i.e. $\varepsilon
\lesssim -2.5\%$.

\textbf{The variation of height, $\widetilde{\Delta
h}=\sqrt{<h^2>-<h>^2}$, in Fig. 1(a) after 10 million MC steps
fluctuates around 0.2~\AA~which is comparable with those found when
using REBO~\cite{EURoJB}. In Figs. 1(b,c) for compressed nanoribbons
of about $\varepsilon$ = -2.0 $\%$ $\widetilde{\Delta h}$ is
0.5~\AA~ after 54 million MC steps. The larger compressive strain
yields a larger height variance.}

Figure~\ref{fig3}(a) shows the variation of the total energy
(Eq.~(\ref{general})) with applied strain at $T$=300\,K.  The
vertical dashed line separates compressive (left) and tensile strain
(right). Square (circular) symbols refer to AC (ZZ) graphene. Notice
that AC and ZZ strained graphene result in the same energy, although
their ripples structure (see Figs.~\ref{fig0}(b,c)) can be rather
different. Note that the energy curve is no longer symmetric around
$\varepsilon=0$ beyond the colored rectangle where
$|\varepsilon|\gtrsim 0.02$. Inside this region the deformation is
symmetric and the harmonic approximation to the total energy works
well as shown by the full black (parabolic) curve in
Fig.~\ref{fig3}(a). The solid curve is a quadratic fit according to
$E_T=E_0+\frac{1}{2}Y\varepsilon^2$ for only positive strains, where
the fitting parameter $Y$ is Young's modulus and $E_0$ is the energy
of the graphene sheet in the absence of strain. We found
$E_0$=0.232$\pm$0.002\,N/m and $Y$=350.42$\pm$3.15\,N/m for  room
temperature. \textbf{The calculated error bars are derived from the
fitting procedure of our numerical  data. The best fit yielded the
smallest deviation from the harmonic behavior.} Our result  for the
 room temperature Young's modulus  is close to the experimental value (340$\pm50$\,N/m)
 and is within the ab-initio results (335-353\,N/m)
and is \textbf{in agreement with}  those obtained from other
classical force fields such as  (LCBOPII~\cite{fasolino} and
Tersoff-Brenner~\cite{TersoffBrenner}) and
Tight-Binding~\cite{poisson}, see Table. II. Note that Perebeinos
and Tersoff  estimated $Y$ at zero temperature and found  1.024
N/$m^2$ (343.04 N/m) ~\cite{Tersoff2009}. Here we calculated $Y$ at
room temperature via stretching-compression simulations.
\textbf{Different force fields are parameterized such that they
describe a set of chosen
  experimental data of particular experimental effects. For example, the VFF model can not
be used to study hydrogenation, melting and defect formation in
either graphene or carbon
 nanotubes sheets, while the REBO has been set-up such that it can be used in those cases.
  The property that the energy can be separated into different energy modes
 and the simplicity of coding the VFF potential are two important advantages of this model.}

Notice that the total energy for AC (square symbols in Fig. \ref{fig3}(a)) and ZZ (circular symbols
in Fig. \ref{fig3}(a)) graphene are almost the same which is in agreement
with the results of Ref.~\cite{poisson}.  Graphene acts
isotropically in the linear elastic limit. Beyond the harmonic regime there is a  small local maximum
 in the energy for compression  which is related to the buckling of graphene.
Notice that in this regime there are small differences between  ZZ
and AC sheets. The buckling threshold is about $\varepsilon_b \simeq
-2.5\%$. The buckling strains is smaller than those found by using
REBO~\cite{neekprb82}, i.e. -0.86$\%$. Notice that both the boundary
conditions and the employed interatomic potentials are responsible
for the difference in the buckling thresholds. The main difference
is due to the different potential. The VFF model is not a bond-order
potential (REBO). As we mentioned in the introduction the bending
modulus predicted by REBO is about
   0.69-0.83 eV which is smaller than the one predicted by the VFF model (2.1 eV). Therefore, we expect a
   larger buckling transition using REBO and a smaller one using VFF model \textbf{(considering the negative sign for
   compressive strains}).
Another important reason for the different result is the
calculations method. Here we used Monte Carlo
  (time is meaningless) and in Ref. [9] we used Molecular dynamics simulations.


Figure~\ref{fig3}(b) shows the contribution of the two important energy terms, i.e.
stretching and bending as given by Eq.~(3) and Eq.~(4),
respectively, for strained AC graphene. Notice that
the stretching energy is larger than the bending and that the rate of increase for stretching
is different.
 In the compression part (i.e the region to the left of the vertical dashed line), after the buckling points these energies are almost constant.
 Thus  increasing compression  beyond the buckling point
does not change the bending and stretching energies.
Figure~\ref{figp} shows the contribution of the different energy
terms (scaled by $E_T$) for three values of the strain for both AC
and ZZ graphene (e.g. the first set of bars to the left refer to
$100\times E_{st}/E_T$ for each particular strain shown in the
legends). From Fig. \ref{figp}, we conclude that the contribution of
the energy terms (Eq.~(4) and Eq.~(7)) \textbf{which are not
determinable}
 in the continuum elasticity energy approach
(Eq.~(\ref{Monge})) are substantial and should be retained when
describing strained graphene\textbf{ at the atomistic scales}

Figure \ref{fig3}(c) shows the variation of the other terms in the energy, Eqs. (5-9), with
strain. The energy for $\pi$-$\pi$ repulsion and out-of-plane increase (decrease) with compression (stretching) and
they  behave opposite to the other terms. In
compression experiments the sheets become strongly corrugated and
neighbor $\pi$ orbitals become more misaligned. In other words the
normal to the adjacent surfaces, e.g. - $\overrightarrow{n_{ijk}}$-
and -$\overrightarrow{n_{ijl}}$- become more misaligned which results in an increase of the total energy.
 Notice also that the bond order term is smaller and negative in the compression part with respect
  to the stretching part. The correlation between the bond
lengths is always negative for the compression part. We found that
the relative contribution of the different energy modes for
stretching and compression of  AC and ZZ graphene are almost the
same, compare Fig.~\ref{figp}(a) and Fig.~\ref{figp}(b).
 However as we see from Figs.~\ref{fig0}(b,c)  the  structure of the ripples
depends on the direction of the applied strain. In the case of AC
the ripples are regular (sinusoidal shape)
while they exhibit an irregular pattern in ZZ graphene. Note that the buckling in plates is generally known to
depend on the plate geometric parameters~\cite{book,neekprb82}. 
Notice that, the dependence of the ripple structure of the graphene sheets on the sheet geometry has been
demonstrated experimentally in Ref.~\cite{arxiv2010}.

\section{Molar heat capacity}
Next, we simulated graphene at different temperatures. Figure \ref{figtemp}(a) shows
the temperature dependence  of the average total energy and
of the six energy terms for $\varepsilon=0$. Notice that  all energy terms vary
linearly with $T$. The quantity $C_{V}=N_AS_0\frac{d\langle E_T\rangle}{dT}$ gives the
potential energy contribution to the molar heat capacity of the system at constant volume, where $N_A$ is
 Avogadro's constant. Note that we first relaxed the volume of the system by performing a
  constant pressure-temperature (NPT) Monte Carlo simulation  (which removes   possible boundary strains). Then we fixed
  the boundaries to the found relaxed size and allowed for  additional thermal relaxation
  (i.e. constant volume-temperature Monte Carlo simulation or NVT).
  During this new thermal relaxation no strain is applied $\varepsilon=0$, thus,
 the calculated heat capacity corresponds to constant volume molar heat capacity.
  Surprisingly, we found that $C_{V}=12.33$\,J\,mol$^{-1}$\,K$^{-1}$ which is almost half of the Dulong-Petit
 value, i.e. $3\Re$=24.94\,J\,mol$^{-1}$\,K$^{-1}$.  Notice that $\langle E_T\rangle$ is the
average of the potential energy of  graphene which is taken over 4
million Monte Carlo steps. Assuming that  the average of the kinetic
term equals the average potential energy ($\langle E_T\rangle$)
according to the equipartition theorem (in the harmonic regime), we
can write
 the total energy $\langle E\rangle=\langle E_T\rangle+\langle K\rangle=2\langle E_T\rangle$ and
  then the total heat capacity is found to
  be $24.66\pm$0.10\,J\,mol$^{-1}$\,K$^{-1}$. The obtained result is close to
   our previous result obtained using REBO, i.e. 24.98$\pm$0.14\,J\,mol$^{-1}$\,K$^{-1}$~\cite{neek-amalPRB2001}.
In Ref. ~\cite{fasolino} the heat capacity at 300 K was found to be
24.2\,J\,mol$^{-1}$\,K$^{-1}$.
 We have performed many simulations at different temperatures for strained graphene and found always a linear
  $\langle E_T\rangle-T$ curves. Fig.~\ref{figtemp}(b) shows the variation of $C_V$ versus strain.
  It is interesting to note that  $C_V$ is slightly lower ($\thicksim 1.0\%$) in  compressed graphene
  as compared to  stretched graphene.\\
\textbf{ Furthermore, we performed several simulations using the
same sample employing the AIREBO potential~\cite{AIREBO} within
LAMMPS software~\cite{lammps}. It is interesting to know that AIREBO
gives (in the range of 10\,K-1000\,K) $C_V$=24.92
J\,mol$^{-1}$\,K$^{-1}$ which was found to be independent of
temperature. Therefore, the VFF model, REBO and AIREBO predicts
temperature independent heat capacity.} \\\textbf{ On the other hand
we found that the VFF model, REBO and AIREBO, give a linear increase
in the carbon-carbon bond length ($a$) with temperature. The
resulting bond length thermal expansion coefficients for the VFF
model, REBO and AIREBO are
$\alpha=\frac{1}{a_0}\frac{da}{dT}=(5.0\pm0.07)\times10^{-6}K^{-1}$,
$(5.0\pm0.03)\times10^{-6}K^{-1}$ and
$(7.0\pm0.04)\times10^{-6}K^{-1}$ respectively. The Gruneisen
parameter is defined as $\gamma=\frac{\alpha B}{C_V \rho}$ where B
is the two dimensional bulk modulus for graphene, i.e $B=12.7$
eV\AA$^{-2}$~\cite{fasolino}, and $\rho$ is the mass density of
graphene, i.e. $\rho=12.0/S_0=7.6\times10^{-4}$g\,m$^{-2}$. Using
our result for $C_{V}$ and $\alpha$ gives $\gamma=0.64$ which is
better estimation for the Gr\"{u}neisen parameter than the one found
in Ref.~\cite{Tersoff2009}, i.e. -0.2, and is closer to the
experimental result, i.e. 2.0~\cite{Gruneisen}.}

\begin{figure*}
\begin{center}
\includegraphics[width=0.45\linewidth]{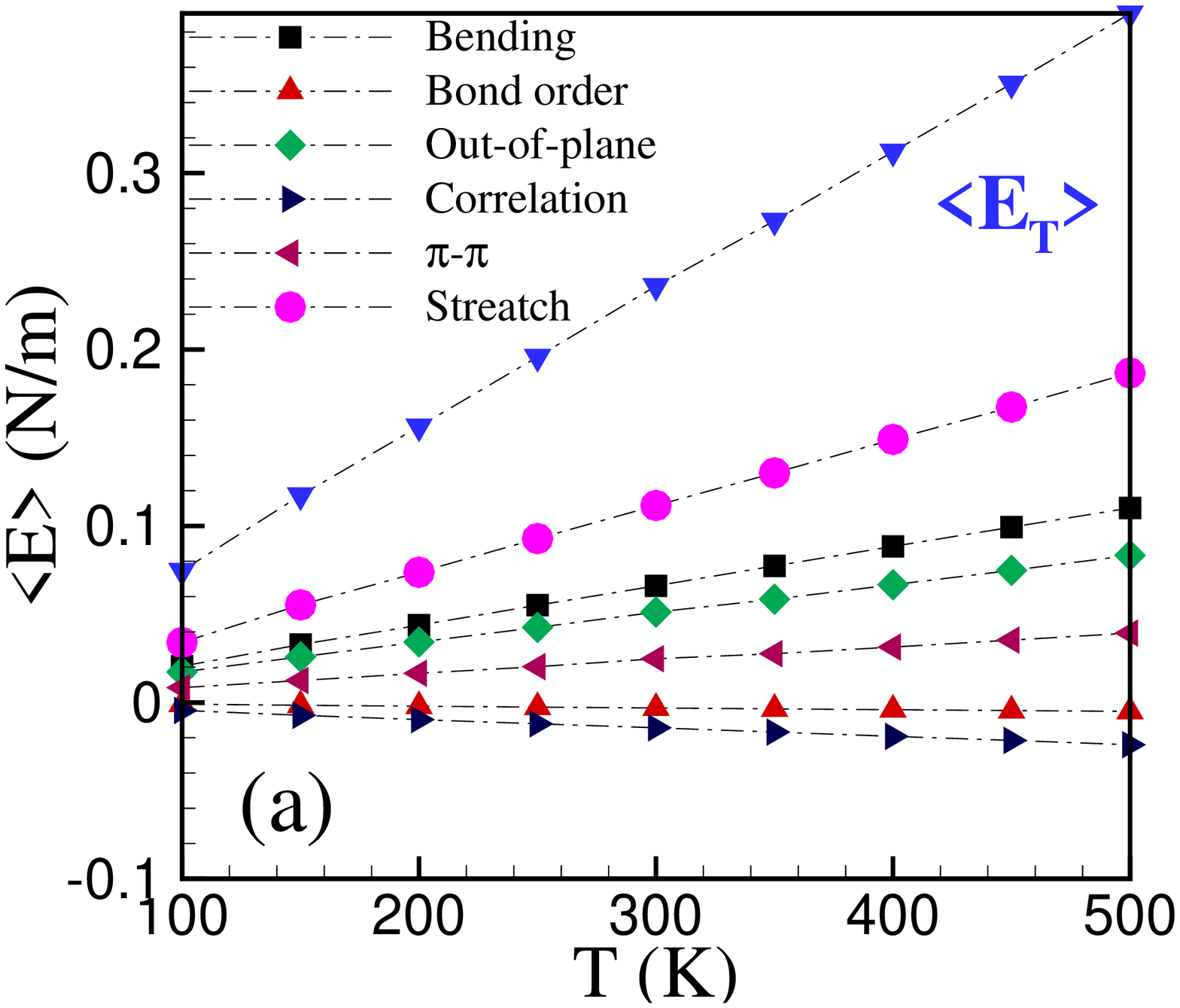}
\includegraphics[width=0.45\linewidth]{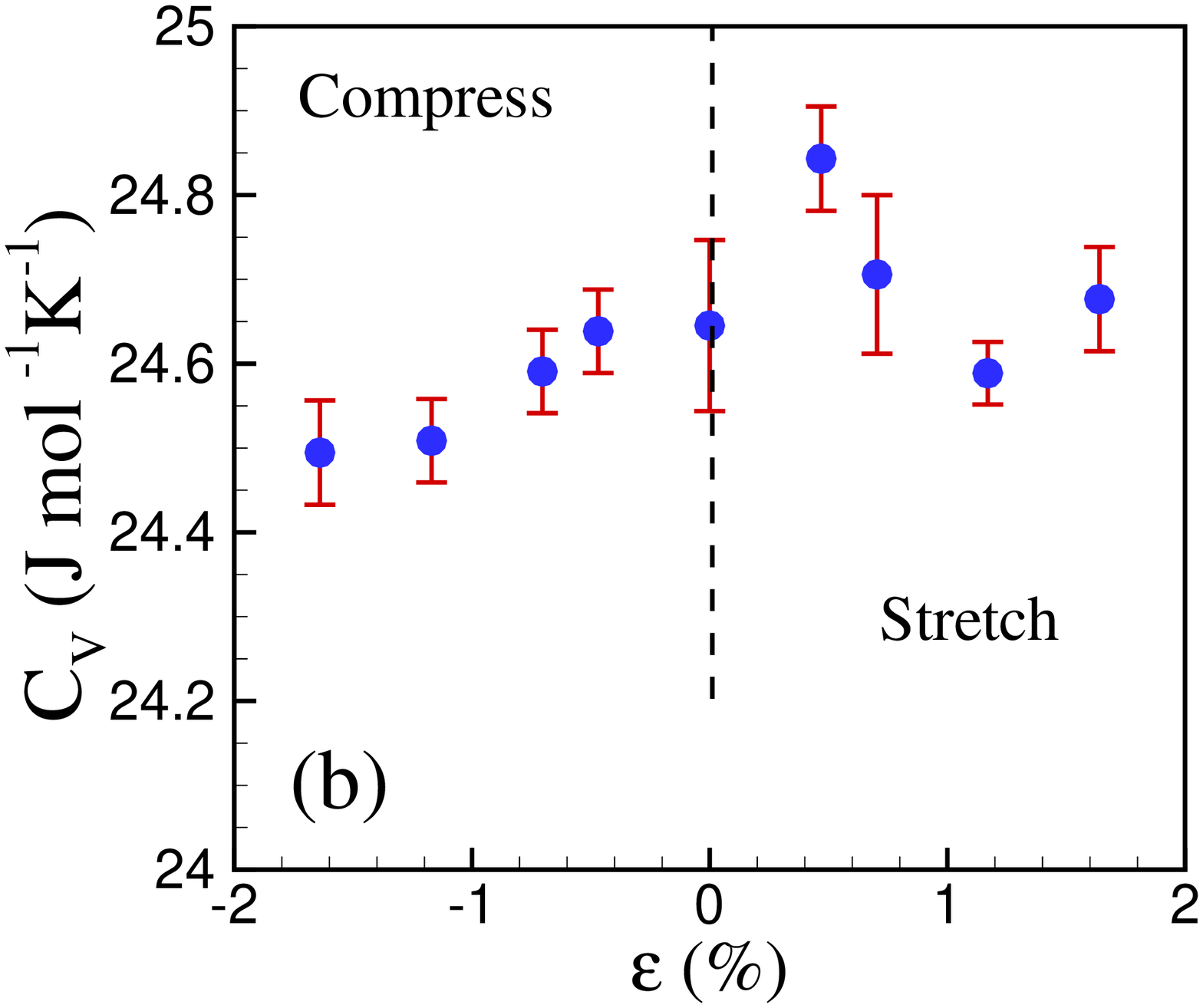}
\caption{(Color online) (a) Temperature dependence of the various
energy modes of a suspended graphene sheet which is suspended along
arm-chair direction. (b) Variation of molar heat capacity at
constant volume for graphene subjected to compressive and tensile
strains. \label{figtemp} }
\end{center}
\end{figure*}

\begin{figure}
\begin{center}
\includegraphics[width=1.0\linewidth]{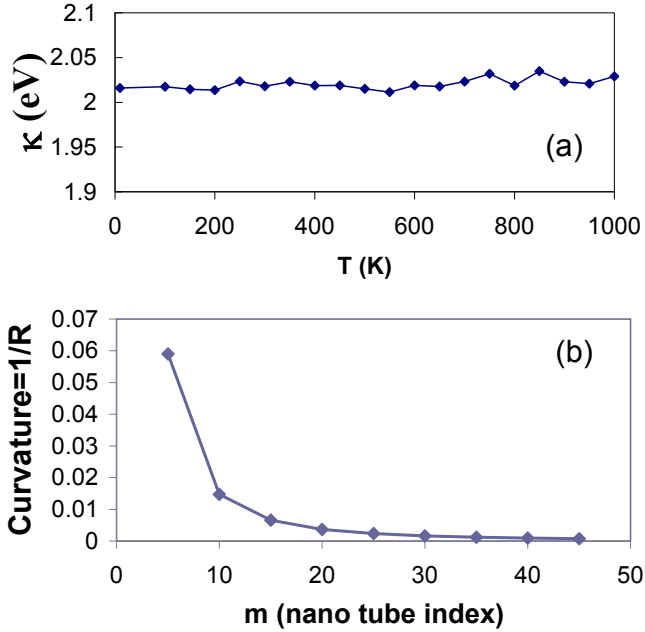}
\caption{(Color online) Temperature dependence of the bending
modulus of graphene (a) and variation of the nanotube curvature
versus nanotube index, i.e $m$. \label{kappa} }
\end{center}
\end{figure}

\section{Temperature effect of the  bending modulus}
A common method for calculating  the bending modulus of
  graphene is by performing several simulations as function of the radius ($R$) of the
   curved tubes and extrapolates the results to $R\rightarrow
   \infty$ (see Fig.~5(b)).  Hence, one can calculate the elastic energy of  carbon nanotubes as a function of the
  inverse square of the radius, $E=\frac{1}{2}\kappa R^{-2}$.
   The coefficient $\kappa$ in the elastic energy gives the bending modulus of
   graphene.
  In order to study the effect of temperature on the bending modulus  of graphene we have carried out several
 NPT Monte Carlo simulations (constant pressure with periodic boundary condition) at different temperatures.
 For each particular temperature we have 8 different tubes.
 In this part of the paper, our systems are different armchair carbon nanotubes with radius
 $R=3\,m^2\,a_0/2\pi$ and initial length 10\,nm.
 We used eight armchair carbon nanotubes with index (m,m) for m=5,10,15,20,25,30,35,40.
 For each particular nanotube with index $m$, we carried out several NPT Monte Carlo
 simulations with periodic boundary condition along the nanotube axis and varying temperatures in the
 range 10 to 1000\,K. Calculating $E_T$ by using Eq.~(2)
 for all nanotubes at temperature $T$, we fitted $\frac{1}{2}\kappa R^{-2}$ to the data and found the bending modulus
 (stiffness), $\kappa$, at $T$. From Fig.~\ref{kappa}(a) we notice that $\kappa$ is practically
\textbf{ temperature independent and is about 2.02\,eV}. Thus the
present VFF model results in a temperature independent bending
modulus. Using membrane theory to calculate the bending rigidity  of
graphene shows that different potentials leads to conflicting
temperature dependence for
 the bending rigidity, e.g. LCBOPII ~\cite{fasolinonature} yields an increase of the
 bending rigidity with temperature while REBO predicts a decreasing dependence~\cite{liuAPL}.

\begin{figure}
\begin{center}
\includegraphics[width=1\linewidth]{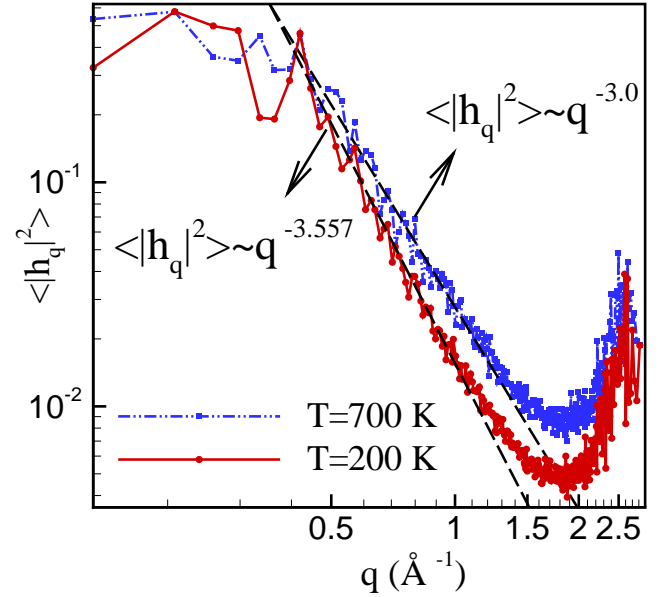}
\caption{(Color online) The absolute value of the
 square of the Fourier transform of atomic heights of C-atoms ($|h_q|^2$) versus the  absolute value
 of the wave vectors of the graphene lattice. \label{Fourier} }
\end{center}
\end{figure}

\section{Scaling properties}
In the harmonic regime the power spectrum of  the graphene
solid membrane  can be obtained by calculating $<|h_q|^2>$ where
 $h_q$ is the Fourier transform of the height of the atoms ($h$) and $q$ is the norm of the
wave vector $\overrightarrow{q}$ $(=(q_x ,q_y)=2\pi
(\frac{n_x}{l_x},\frac{n_y}{l_y}))$ with integers $n_x$ and $n_y$
where $l_x$ and $l_y$ are the longitudinal and lateral size of the
graphene sample. It is important to note that in this section we
simulated a graphene sheet with initial size $l_x=230.04\,\AA$~ and
$l_y=221.35\,\AA$~($N$=19440) using standard  NPT Monte Carlo
simulations  with periodic boundary conditions in both directions
(the method is similar to that reported in Refs. [4,10]). We
estimated the spectral modes $h_q$ by fitting $|h_q|^2$ to a
$q^\alpha$ function,
 from which we  extract the power law $\alpha$. Figure~\ref{Fourier} shows  the
 variation of $|h_q|^2$ (averaged over 500 Monte Carlo realizations where 5 neighboring points were accumulated and
 averaged to a single point in order to make the curves smoother) versus $q$ for graphene
 at two temperatures 200 K and 700 K.
 The dashed lines are power law fits.
Notice that $\alpha \neq -4$ which clearly indicates that
 anharmonic effects are present in the used VFF potential.
 Moreover we see that $\alpha$ decreases with decreasing
  temperature ($\alpha=$-3.0 for T=700\,K and $\alpha=$-3.557 for T=200\,K)
  which hints that a more  harmonic behavior is found at
   low temperature when  using the VFF potential. The latter temperature
   dependence is in agreement with the  REBO predictions~\cite{EURoJB}.
    Notice that the REBO is a bond-order interatomic potential.
   Note that the peaks in Fig.~\ref{Fourier} are related to first Bragg-peak, $4\pi/3a_0=2.94\AA$~ due to the
  discreteness of the graphene lattice.
\textbf{Notice that the modulation amplitude in  Fig. 6 for T=200 K
is about 0.5\AA~and for T=700 K is about 0.7\AA~ which are
temperature and size dependent quantities, the larger the size the
larger the amplitudes (here graphene has dimension $221\times230$
\AA$^2$). Here, we did not study the effect of size and refer the
reader to Refs.~\cite{fasolinonature,fasolino} where such a study
can be found.}


\section{Conclusions}
In this study, we showed that the recently proposed \emph{valence
force field} model (VFF)~\cite{Tersoff2009} for graphene enables to
compare the contribution of the different energy terms when
straining graphene. In a stretching experiment the main energy
contributions are due to stretching and bending terms while for
compressive strains also other terms such as out-of-plane and
$\pi-\pi$ interaction terms play an important role. We found that
using such a classical approach gives  accurate values for the
Young's modulus at room temperature which are found to be as
accurate as those using ab-initio methods. The calculated Young's
modulus is close to the experimental result.  The total energy is
quadratic in $\varepsilon$  for strains smaller than $|2\%|$. The
current VFF model predicts a  temperature independence
 bending modulus. The temperature dependence   of the total strain energy yields an acceptable value for the molar
  heat capacity of graphene which is too a large extend independent of the applied strain.
The Gruneisen parameter is found to be positive and about 0.64.

~~~~

\emph{{\textbf{Acknowledgment}}}. We acknowledge helpful comments by
V. Perebeinos, S. Costamagna, A. Fasolino and J. H. Los. This work
was supported by the Flemish science foundation (FWO-Vl) and the
Belgium Science Policy~(IAP).

~~~

\end{document}